\documentclass[prl,twocolumn,floatfix,amsfonts,letterpaper]{revtex4}
 
\usepackage{amsfonts}
\usepackage{graphicx,graphics,epsfig,epstopdf}
\usepackage{dcolumn}
\usepackage{float}
\usepackage{bm}
\usepackage[usenames,dvipsnames]{color}
\usepackage{color}

\newcommand{\com}[1]{}

\newcommand{\bS}{\mathbf{S}}

\newcommand{\rr}{\mathbf{r}}

\newcommand{\bsigma}{{\boldsymbol\sigma}}

\newcommand{\z}{{\hat z}}
\newcommand{\ve}{{\varepsilon}}
\newcommand{\up}{{\uparrow}}

\newcommand{\beq}{\begin{equation}}
\newcommand{\eeq}{\end{equation}}
\newcommand{\beqa}{\begin{eqnarray}}
\newcommand{\eeqa}{\end{eqnarray}}

\begin{document}

\title{Fractionalization in spontaneous integer quantum Hall systems}

\author{ Rodrigo A. Muniz$^{1,2}$, Armin Rahmani$^1$ and Ivar Martin$^1$ } 
\affiliation{$^1$Theoretical Division, Los Alamos National Laboratory, Los Alamos, NM 87545, USA \\ 
$^2$International Institute of Physics - UFRN, Natal, RN  59078-400, Brazil}

\begin{abstract}

It is widely believed that integer quantum Hall systems do not have fractional excitations. Here we show the converse to be true for a class of systems where integer quantum Hall effect emerges spontaneously due to the interplay of itinerant electrons and noncoplanar magnetic ordering. We show that magnetic $Z_2$ vortices in such systems, whose topological stability is guaranteed by the structure of the order parameter, indeed carry fractional charge.
\end{abstract}

\maketitle

The discoveries of the integer and fractional Quantum Hall (QH) effects have revolutionized condensed matter physics: the important concept of a topological invariant was introduced to explain the quantized Hall conductivity of the former~\cite{TKNN}, while the novel notion of topological order, i.e., a type of nonlocal order with no Landau symmetry-breaking and no local order parameter, was introduced to describe the latter~\cite{Wen90}. Topological order goes hand in hand with exotic phenomena such as fractional charge and statistics~\cite{Wen90}. While the strongly-correlated, topologically-ordered fractional QH systems indeed have fractional quasi-particles \cite{Laughlin, Moore-Read}, their more traditional integer counterparts are assumed to only support natural excitations with integer charge. Contrary to this assumption, we predict in this paper that factional excitations can generically emerge in a class of integer QH systems.

The existence of fractional excitations in a fractional QH liquid follows from the Laughlin's argument: upon adiabatic local ``insertion" of a flux quantum, a fractional charge $q = \sigma_{xy} e$, where $\sigma_{xy}$ is the fractional Hall conductivity, flows in from infinity. Since integer flux quantum can be ``gauged away,'' the charge $q$ is in fact the charge of an elementary quasiparticle. Naturally, insertion of a fractional flux can also lead to the same result with an integer Hall response. Fractional fluxes, however, cannot be gauged away, and do not naturally appear in  a traditional integer QH system, i.e., two-dimensional electron gas in a magnetic field. (In principle, half-vortices could be introduced in artificial heterostructures, by placing a QH system in proximity to a superconductor \cite{Weeks}.) It may thus appear that the presence of intrinsic fractional excitations in a QH system requires a fractional $\sigma_{xy}$.

However, besides topological order, charge fractionalization can also occur through another mechanism: coupling of noninteracting electrons to topologically stable defects in ordered backgrounds, such as a pattern of lattice distortions in polyacetylene or graphene-like structures \cite{Jackiw,ssh,hou, chamon} or in a superconducting vortex \cite{JR, volovik, Read-Green, Ivanov}. The presence of an ordered background requires spontaneous symmetry breaking, which does not happen in traditional integer QH systems. But QH effect can also emerge as a result of spontaneous time-reversal symmetry breaking in the absence of a magnetic field \cite{ander, Haldane88, nagaosa, shindou, MB2008, raghu2008}. As we will show in this paper, such spontaneous integer QH systems can support topologically stable defects, which remarkably act as an effective fractional flux, thus giving rise to natural excitations with fractional charge.

\begin{figure}[ht]
\begin{tabular}{r}
\includegraphics[width=.87\columnwidth]{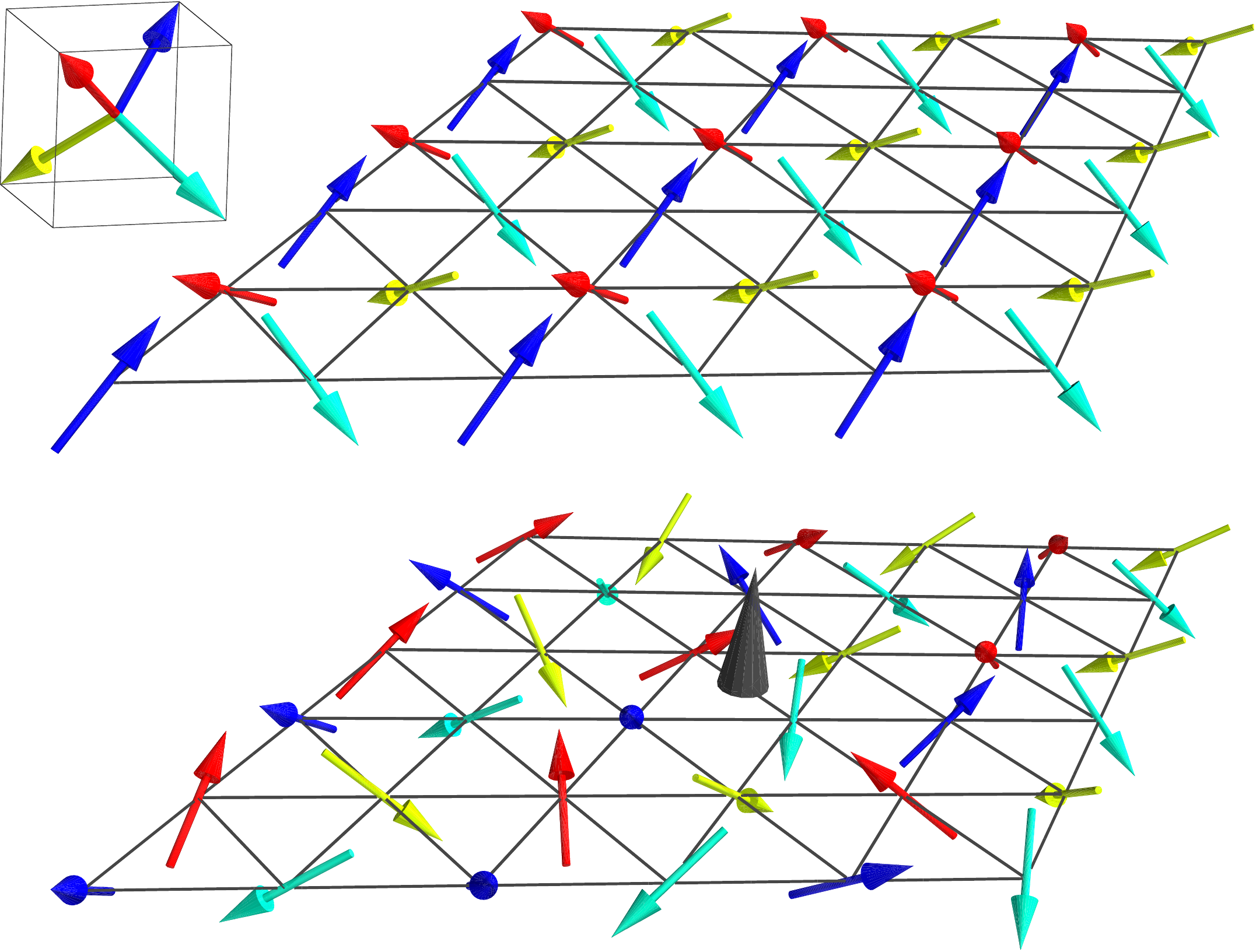}
\end{tabular}
      \put(-233,83){ (a) } 
      \put(-233,-13){ (b) }
\caption[]{(color online). (a) Chiral spin ordering on a triangular lattice. Four orientations of the local
magnetic moments correspond to the normals to the sides of a regular tetrahedron.
(b) Vortex configuration.}
\label{FIG:1}
\end{figure}

Spontaneous QH effect can be induced by noncoplanar magnetic ordering, 
as depicted in Fig.~\ref{FIG:1}a, serving as the background for itinerant electrons \cite{nagaosa, shindou, MB2008}. 
Even though in this work we focus on the specific model of Ref.~\cite{MB2008} on the triangular lattice, we would like to emphasize that our qualitative conclusions are independent of the microscopic details, and rely on only two ingredients: noncoplanar magnetic ordering
and the presence of an incompressible integer QH electronic state. We can, for instance, apply the same arguments to the model of Ref.~\cite{nagaosa} on the kagome lattice, as shown in the supplementary material~\cite{supp}.

In the model of Ref. \cite{MB2008} (Fig.~\ref{FIG:1}a), the magnetic ordering spontaneously and completely breaks $O(3)$ symmetry.  
The existence of topologically stable vortex defects in a noncoplanar magnetically ordered medium follows from the nontrivial fundamental homotopy group of the order parameter space, $\pi_1(O(3)) = \pi_1(SO(3)) = Z_2$ ~\cite{Kawamura,supp}. An example of the vortex spin texture is shown in  Fig. 1b. 
For instance, it can be obtained by rotating the order parameter (every magnetic moment)  by a position-dependent angle 
\beq \label {eq:phi}
\phi(\rr) = \nu   \arg (x + iy)
\eeq
around the $\z$ axis (assuming the vortex core is at the origin); $\nu$ is the winding of the vortex. The two topological classes are set by the parity of $\nu$. Unless stated otherwise, here we consider the $\nu = 1$ case.

A model that gives rise to the spontaneous QH state is the following Kondo lattice model:
\beq
H =  -t_{ij} c_{i\alpha}^\dag c_{j\alpha}-\mu c_{i\alpha}^\dag c_{i\alpha}+ J   c_{i\alpha}^\dag \bS_i\cdot\bsigma_{\alpha\beta} c_{i\beta}\label{eq:H},
\eeq
where electrons hop on a two-dimensional (2D) lattice and interact with the local magnetic moments via onsite exchange interaction. Summation over repeated site (roman) and spin (greek) indices is implied. Here $t_{i,j}$ is the intersite hopping, $J$ is the exchange interaction constant, $\bS_i$ is the local magnetic moment (assumed classical), $\bsigma = (\sigma^x, \sigma^y, \sigma^z)$ is the vector of Pauli matrices, and $c_{i\alpha}$ is the operator of electron annihilation on site $i$ with spin $\alpha$. 
On a triangular lattice, the model has a spontaneous QH groundstate (Fig. 1a.) for several electron densities \cite{MB2008,Akagi2010,Kumar2010, Kato2010}.

The magnetic texture may become inhomogeneous either due to thermal or quantum fluctuations, or due to the presence of topologically stable defects, such as vortices (Fig. 1b). Such inhomogeneous states can be mapped onto a state with homogeneous order parameter, but in the presence of an effective (in general) non-Abelian gauge potential. This mapping allows one to conveniently calculate the charge and spin currents in response to the order parameter distortions. In particular, the vortex configuration corresponds to a spatially localized non-Abelian flux, which can be used to determine the charge of the vortex. Indeed, an inhomogeneity corresponds to a rigid position- and possibly time-dependent distortion of some reference state, $\bS_i = {\cal R}(\rr_i, t)\bS_i^0$.
The rotation of the order parameter in the classical spin space can be transformed into a unitary rotation $U(\rr_i,t) \equiv U_i$ of the electron spinors, according to $U_i^\dag \bsigma\cdot\bS_i U_i = \bsigma\cdot\bS_i^0$. Introducing new fermions $\psi_i = U(\rr_i, t)^\dag c_i$ (spin index is suppressed), the Hamiltonian (\ref{eq:H}) becomes
\beqa
H = &- i \psi_{i}^\dag U_i^\dag \partial_t U \psi_{i}  -t_{ij} \psi_{i}^\dag  U_i^\dag U_j \psi_{j}\nonumber\\
&-\mu \psi_{i}^\dag  \psi_{i}+ J  \bS_i^0\cdot \psi_{i}^\dag \bsigma \psi_{i}\label{eq:Hrot}.
\eeqa

Assuming that the variation of the texture is slow on the lattice constant scale, we can make an expansion,
$U_i^\dag U_j = U_i^\dag [U_i + (\rr_j - \rr_i)\cdot \nabla U_i]$. 
It is convenient to introduce $SU(2)$ vector potential ${\bf A^{\nu}} = - i U^\dag \partial^\nu U \equiv {\cal A}_a^\nu \sigma_a$, with the indices $\nu = \{t,x,y\}$ representing the space-time components and $ a = \{1,2,3\}$ the $SU(2)$ generators. The Hamiltonian can then be written as 
\beq
H = H_0 - J_a^\nu {\cal A}_a^\nu, \label{eq:HJ}
\eeq
where $H_0$ is the Hamiltonian corresponding to the static uniform spin structure, and the currents are defined as
\beqa
J_a^0& =&\psi(\rr_i,t)^\dag \sigma_a\psi(\rr_i,t),\\ 
J_a^x &= & i t_{ij}(x_i - x_j) \psi(\rr_i,t)^\dag \sigma_a\psi(\rr_j,t), \\
J_a^y &= & i t_{ij}(y_i - y_j) \psi(\rr_i,t)^\dag \sigma_a\psi(\rr_j,t).
\eeqa
Denoting the $2\times 2$ unit matrix by $\sigma_0$, for $a=0$ the definitions above also give the physical charge density and current operators 
\footnote{The physical observables are defined in terms of the original $c$-fermions, and while charge density is the same in the old and the new bases, $c_i^\dag c_i = \psi(\rr_i)^\dag\psi(\rr_i)$, other operators, such as spin, need not be.}.

If the texture is slowly  varying in time (compared with the inverse energy gap in the spectrum), as well as in space, the vector potential ${\cal A}_a^\nu$ is small and the current expectation values can be calculated with the linear-response theory. Through an explicit calculation for the model of Ref. \cite{MB2008}, we find the following nonvanishing responses~\cite{supp}:
\beq\label{eq:sigma}
\sigma^{00}_{xy} = -e^2/h, \qquad\sigma^{aa}_{xy} =  e^2/3h\  ({\rm for\ } a\ne 0),
\eeq
which determine the charge and spin currents through $\langle J_a^\eta \rangle= {\sigma^{aa}_{xy}} \epsilon_{\eta\mu\nu}\partial^\mu {\cal A}_a^\nu$.  
The signs of both conductivities are flipped by switching between $^3/_4$ and $^1/_4$ fillings, or by changing the sign of the chiral ordering.

The vortex texture is necessarily singular near the core; therefore, the gauge transformation that one needs to apply to unwind it will be singular as well. The simplest transformation that takes the vortex texture into a uniform one is  $e^{i\sigma_3 \phi(\rr)/2}$, where $\phi$ is the angle of rotation~(\ref{eq:phi}) around the $z$ axis.
However, since upon going around the vortex, $\phi(\rr)\to \phi(\rr) + 2\pi$, the unitary changes sign, this would correspond to antiperiodic boundary conditions for fermions along a line connecting the vortex to infinity. 
To avoid this complication,  the above $SU(2)$ transformation can be augmented by a $U(1)$ one \cite{WA}. The combined transformation $U(\rr)= e^{i(\sigma_3+1) \phi(\rr)/2}$ is only acting on up-fermions. Its associated gauge potential is
$${\bf A^\nu} =\frac{1+ \sigma_3}{2}\,\partial^\nu \phi,$$
which has the field strength zero everywhere except for the vortex core.

Due to the singular nature of the gauge potential, we cannot directly apply the linear-response formalism in the vicinity of the vortex core. To calculate the vortex quantum numbers, we can instead invoke an analogue of the Laughlin argument. The flux of the non-Abelian gauge field through the vortex core is 
$${\bf \Phi} = \oint {\bf A} d\rr = ({1+ \sigma_3})\pi.$$
Now, suppose that the flux is turned on adiabatically from zero to $\bf \Phi$. That will generate a non-Abelian emf acting on electrons, which at large distances from the core will be nearly uniform (tangential to any circle centered  at the vortex). The vortex quantum numbers are then obtained by integrating the associated currents generated in response to this emf.

Before proceeding, let us note that the Laughlin argument relies on two conditions. First, we need to have a continuous sequence of \textit{gapped} Hamiltonians connecting the one with flux zero to the one with $\bf \Phi$. Second, we need a continuity equation relating the currents we can calculate in linear response to quantum number densities. We have explicitly identified a sequence of gapped Hamiltonians in the limit of large $J$ and we thus expect that an adiabatic process exists for arbitrary $J$. The quantum numbers of interest for us are charge and spin. Since total electron number commutes with the Hamiltonian, the charge current strictly satisfies a continuity equation. We do not have such continuity equation for the spin current, and thus our vortex defects do not have a well-defined spin quantum number.

 However, since the divergence of the induced current is zero far from the vortex, it is expected that the spin current will be nearly conserved, except near the vortex where it accumulates. This suggests that the \textit{magnetization} attached to a vortex might be still close to the expected value obtained from integrating the spin current. With the assumptions above, since the final flux both in spin-$\sigma_3$ and charge channels is half of the flux quantum, the accumulated charge and magnetization should be
\footnote{Since the gauge transformation only involves $\sigma_3$ Pauli matrix, the spin density in this channel is identical both in $c$ and $\psi$ bases.}
\beq
q = \sigma^{00}_{xy}/2,\qquad
m_z \approx \sigma^{33}_{xy}/4. \label{eq:s_z} \label{eq:qsz}
\eeq
(the extra $^1/_2$ in the expression for $m_z$ is due to the fact that electron spin is $\bsigma/2$.)
For less symmetric systems such as in Ref.~\cite{nagaosa} there are contributions from mixed conductivities, like $\sigma^{03}_{xy}$ or $\sigma^{13}_{xy}$ for instance, to $q$ and $m_z$ as well as to other spin components.

There are also other possible choices of the gauge transformation that unwinds the vortex, e.g., with the gauge potential ${\bf A^\nu} =({-1+ \sigma_3})\,\partial^\nu \phi/2$.  While such different choice does not change the magnetization, the charge accumulated in the vortex core changes sign. This ambiguity is naturally understood in terms of the electron occupancy of a particular localized electronic state, $\ve_0$, inside the spectral gap. When this state is empty the charge of the system is $-\,^1/_2$ and when it is occupied it is $+\,^1/_2$, all relative to the uniform state. In general, there can be more than one localized state inside the vortex core. Occupying any of these states increases the vortex charge by one electron charge [this corresponds to more general gauge choices, ${\bf A^\nu} =({1 + 2n + \sigma_3})\,\partial^\nu \phi/2$, with $n$ any integer]. However, this cannot change the fact that the charge of the vortex has to remain half-odd-integer for an odd vorticity.  On the other hand, for an even vorticity, the charge induced according to the Laughlin argument will be integer. This is consistent with the homotopy classification that says that double vortex can be  smoothly connected to a uniform state.

\begin{figure}[ht]
\includegraphics[width=8 cm]{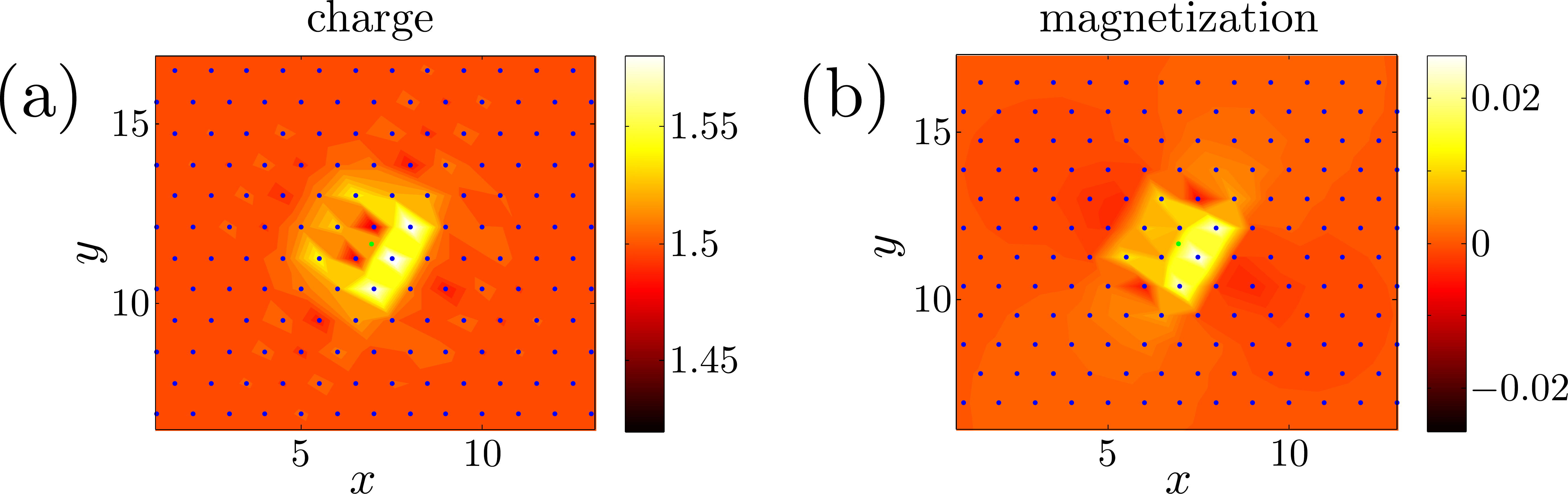}
\caption[]{(color online). Charge (a) and magnetization (b) density distributions around a vortex. The vortex is indicated by the green dot at the center. The system parameters are as follows: $J \to \infty$, $L =30$, $h=0$ at $^3/_4$ electronic filling.
} 
\label{FIG:2}
\end{figure}

\begin{figure}[ht]
\includegraphics[width=8 cm]{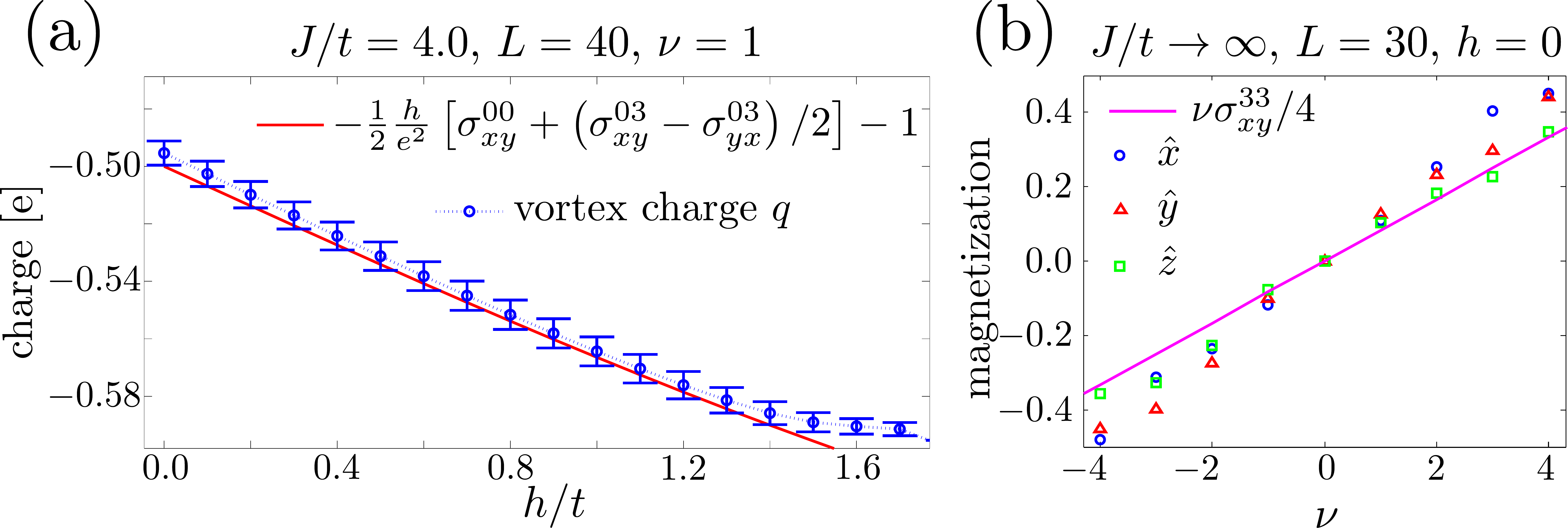}
\caption[]{(color online).
(a) Zeeman field dependence of the vortex charge. Blue circles were obtained by exact diagonalization. The red solid line is the expected result from a Laughlin adiabatic argument using the uniform dc Hall conductivities of an infinite uniform system shown in Fig. 1a. (b) Net magnetization accumulated nearby the vortex for different windings and with rotations around the 3 axes. The solid line corresponds to Eq.~(\ref{eq:s_z}). Both results were obtained for $^3/_4$ electronic filling.}
\label{FIG:3}
\end{figure}

We now check the above results numerically using the explicit model of Ref. \cite{MB2008} (Fig. 1).  We plot, in Fig.~\ref{FIG:2}, the charge and magnetization distribution in the vicinity of the vortex core.
As expected, the charge localized in the core is half-odd-integer for odd winding and integer for even winding. 
The agreement between the vortex magnetization obtained numerically with what is expected from Eq.~(\ref{eq:s_z}) is not perfect because the spin current is not conserved. 
Nevertheless, particularly for large $J$, the discrepancy is not too large and we verify that the vortex spin polarization approximately scales with the vortex winding, as shown in Fig.~\ref{FIG:3}b.

We have also considered deviations from the fully symmetric assumptions. In particular, we added a Zeeman field  $h$ along $\z$ axis acting on electrons, i.e $H\rightarrow H+h\sum_i    c_{i\alpha}^\dag \sigma^z_{\alpha\beta} c_{i\beta}$. We verified that as long as the spectral gap does not close, the charge Hall conductivity does not change. However, in contrast to the fully symmetric case, new nonzero response functions emerge, namely $\sigma^{03}_{xy}, \,\sigma^{03}_{yx} \ne 0$, which correspond to the charge response to ${\cal A}_3$, the $\sigma_3$ component of the gauge potential.
 In Fig.~\ref{FIG:3}a, we plot the dependence of the vortex charge on $h$. The solid line is $q = -\sigma_{xy}/2 - (\sigma^{03}_{xy} - \sigma^{03}_{yx})/4-1$, which directly follows from the application of the Laughlin argument. Note that the charge is generally irrational and determined modulo an integer. The agreement is very good, all the way to the value of $h$ where the gap in the electronic spectrum closes. 
 
The energetics of $Z_2$ vortices is similar to the usual $Z$ vortices for $SO(2)$ order parameter~\cite{Kawamura}: i.e., the energy of an isolated vortex scales logarithmically with the system size. Nevertheless, pairs of log-confined vortex pairs will appear due to thermal fluctuations at finite temperatures. 
In addition, inclusion of quantum spin dynamics may lead to an intriguing possibility that the {\em quantum} fluctuations transform the noncoplanar ordered state into a chiral spin liquid~\cite{KL, WWZ, wiegmann} at zero temperature. There, the fractionally charged $Z_2$ vortices discussed above may turn into deconfined point-like excitations.

Finally, we comment on the vortex exchange statistics. There are two distinct possibilities for the case of two vortices: (1) the total charge of two vortices  is even, and (2) the total charge is odd.  The former case will be realized if the vortices are pulled apart from the uniform ``vacuum": since the initial state has total charge zero (relative to the uniform background), the state with two vortices will keep the same charge. Since the charge of an individual vortex is half-odd-integer, for large inter-vortex separation there are two energetically equivalent ground states that correspond to vortex charge configurations $(\,^1/_2, -\,^1/_2)$
and $(-\,^1/_2, \,^1/_2)$.  
This degeneracy can lead to non-Abelian exchange statistics, but unlike the Majorana states in the p-wave superconductor-like systems \cite{Read-Green, Ivanov}, there is no topological protection. In other words, any local disorder can shift the bound state energy in a given vortex and lift the degeneracy.

The case (2) can be obtained, e.g., upon electron or hole topological doping of the system with equally (half-odd-integer) charged vortices.
In this case, there is no ground-state degeneracy and exchanging two vortices can only give rise to an Abelian Berry phase. The anyonic character of the vortices follows from a simple argument. 
Consider two indistinguishable distant vortices with single winding and equal charge.
Near each vortex the gauge potential is ${\bf A^\nu} =({1+ \sigma_3})\,\partial^\nu \phi/2$, with $\phi$ being the angular coordinate relative to the given vortex core. Therefore, each vortex has a non-Abelian gauge flux attached to it, equivalent to a $U(1)$ flux $2\pi$ in the spin-up channel. The number of up electrons occupying each vortex is $n_\up = (q + 2m_z)/2$.  The statistical angle due to the flux attachment is the product of the charge and half of the flux, which for the idealized case of Eq. (\ref{eq:qsz}) is  $\theta = \pi (\sigma^{00}_{xy} + \sigma_{xy}^{33})/4$. Note that the deviations of $m_z$ from this idealized case are small for $J\rightarrow\infty$ (see $\nu=1$ data in Fig.~\ref{FIG:3}b). The numerical observation of the statistical angle, however, is complicated by the spin nonconservation.

In summary, we have shown that spontaneous integer quantum Hall systems, which emerge from the interplay of itinerant electrons and noncoplanar magnetic ordering, generically support topologically stable excitations with fractional charge and anyonic statistics. Promising candidates materials which may exhibit this physics could be the systems of Na$_x$CoO$_2$ type, which near $x = 0.5$ are known to have a noncollinear order, as well as anomalously large Hall response \cite{cava}. The fractional charge predicted in this work may be accessible through direct imaging of the local charge profile, as shown in Fig.~\ref{FIG:2}, e.g., by scanning force microscopy. Also anyonic exchange statistics may have unusual consequences in real materials. Perhaps the most intriguing among them is the possibility of the Anyonic Superconductivity \cite{LaughSC, wilczek, wiegmann}.
When a system is doped away from the chiral Mott insulating state, it may energetically prefer to  accommodate the carriers by creating vortices with intragap states.  As we have just argued, such occupied vortex states are anyons, which, at finite density and low enough temperature, may go into a superconducting state \cite{wilczek}.

{\acknowledgements We thank  C. Batista, C. Chamon, C.-Y. Hou, D. Ivanov, A. Morpurgo, D. Podolsky, S. Ryu, S. Sachdev and L. Santos for helpful discussions.  This work was carried out under the auspices of the National Nuclear Security Administration of the U.S. Department of Energy at Los Alamos National Laboratory under Contract No. DE-AC52-06NA25396 and supported by the LANL/LDRD Program. R. Muniz also thanks CNPq (Brazil) for financial support. }



\begin{thebibliography}{10}
\expandafter\ifx\csname url\endcsname\relax
  \def\url#1{\texttt{#1}}\fi
\expandafter\ifx\csname urlprefix\endcsname\relax\def\urlprefix{URL }\fi
\providecommand{\bibinfo}[2]{#2}
\providecommand{\eprint}[2][]{\url{#2}}





\bibitem{TKNN}D. J. Thouless, M. Kohmoto, M. P. Nightingale, and M. den Nijs, Phys. Rev. Lett. \textbf{49}, 405 (1982). 

\bibitem{Wen90} X. G. Wen, Int. J. Mod. Phys. B \textbf{4}, 239 -271 (1990).

\bibitem{Laughlin} R. B. Laughlin, Phys. Rev. Lett. {\bf 50}, 1395-1398 (1983).

\bibitem{Moore-Read} G. Moore and N. Read, Nucl. Phys. B {\bf 360}, 362-396 (1991).

\bibitem{Weeks} C. Weeks, G. Rosenberg, B. Seradjeh and M. Franz, Nature Physics \textbf{3}, 796 (2007).


\bibitem{Jackiw} R. Jackiw and C. Rebbi, Phys. Rev. D \textbf{13}, 3398 (1976).

\bibitem{ssh} W. P. Su, J. R. Schrieffer, and A. J. Heeger
Phys. Rev. Lett. {\bf 42}, 1698 (1979).  

\bibitem{hou} C.-Y. Hou, C. Chamon, and C. Mudry, Phys. Rev. Lett. \textbf{98}, 186809 (2007).



\bibitem{chamon} C. Chamon {\em et al.}, Phys. Rev. B {\bf 77}, 235431 (2008).

\bibitem{JR}  R. Jackiw and P. Rossi,  Nucl. Phys. B {\bf 190}, 681 (1981).

\bibitem{volovik} G. E. Volovik and V. M. Yakovenko, J. Phys.: Cond. Mat {\bf 1},  5263 (1989).

\bibitem{Read-Green} N. Read and D. Green, Phys. Rev. B {\bf 61}, 10267 (2000).

\bibitem{Ivanov} D. A. Ivanov, Phys. Rev. Lett. {\bf 86}, 268 (2001).



\bibitem{ander} P. W. Anderson and H. Hasegawa, Phys. Rev. {\bf 100}, 675 (1955).

\bibitem{Haldane88} F. D. M. Haldane, Phys. Rev. Lett. {\bf 61}, 2015 (1988).

\bibitem{nagaosa} K. Ohgushi, S. Murakami, and N. Nagaosa, Phys. Rev.
B {\bf 62}, R6065 (2000).



\bibitem{shindou} R. Shindou and N. Nagaosa,
Phys. Rev. Lett. {\bf 87}, 116801 (2001).

\bibitem{MB2008} I. Martin and C. D. Batista, Phys. Rev. Lett. {\bf 101}, 156402 (2008).

\bibitem{supp} See supplemental material for details.

\bibitem{raghu2008}S. Raghu, X.-L. Qi, C. Honerkamp, and  S.C. Zhang, Phys. Rev. Lett. {\bf 100}, 156401 (2008).

\bibitem{Akagi2010} Y. Akagi and Y. Motome, J. Phys. Soc. Jpn. \textit{79} 083711 (2010). 
%
\bibitem{Kumar2010} S. Kumar and J. van den Brink, Phys. Rev. Lett. \textbf{105}, 216405 (2010).
%
\bibitem{Kato2010}Y. Kato, I. Martin, and C. D. Batista, Phys. Rev. Lett. \textbf{105}, 266405 (2010).




\bibitem{Kawamura} H. Kawamura and S. Miyashita, J. Phys. Soc. Jpn, \textbf{53}, 4138 (1984).
\bibitem{WA} A. G. Abanov and P. B. Wiegmann, Nucl. Phys. B {\bf 570}, 685 (2000).

\bibitem{KL} V. Kalmeyer and R. B. Laughlin, Phys. Rev. Lett. \textbf{59}, 2095 (1987).
 \bibitem{WWZ} X. G. Wen, F. Wilczek, and A. Zee, Phys. Rev. B \textbf{39}, 11413 (1989). 
\bibitem{wiegmann} P. Wiegmann, Prog. Theor. Phys. Suppl. {\bf 107}, 243 (1992).

\bibitem{LaughSC} R. B. Laughlin, Phys. Rev. Lett. {\bf 60}, 2677 (1988).

\bibitem{wilczek} F. Wilczek, {\em Fractional Statistics and Anyon Superconductivity} (World Scientific, Singapore, 1990).


\bibitem{cava} M. L. Foo {\em et. al.}, Phys. Rev. Lett. {\bf 92}, 247001 (2004).

\end{thebibliography}
\end{document}


\title{Supplemental Material for ``Fractionalization in spontaneous integer quantum Hall systems''}
\author{ Rodrigo A. Muniz$^{1,2}$, Armin Rahmani$^1$ and Ivar Martin$^1$ } 
\affiliation{$^1$Theoretical Division, Los Alamos National Laboratory, Los Alamos, NM 87545, USA \\ 
$^2$International Institute of Physics - UFRN, Natal, RN 59078-400, Brazil}

\maketitle

\section{$SO(3)$ vortices}

\begin{figure}[ht]
\includegraphics[width=9cm]{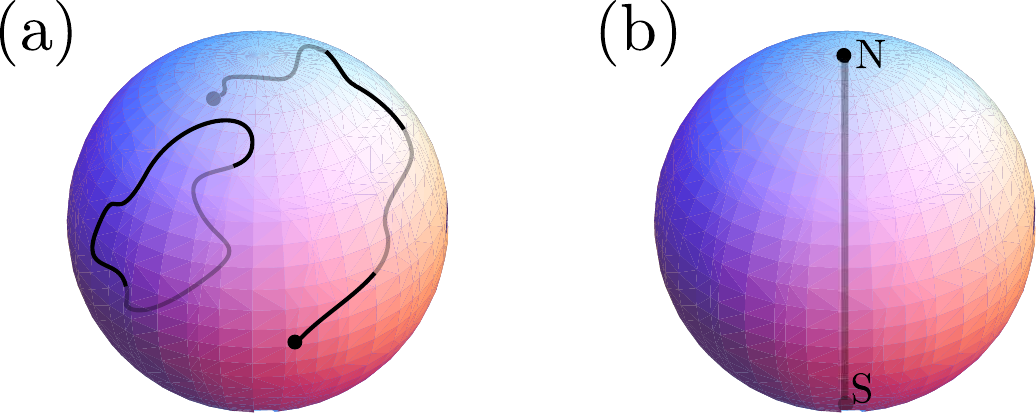}
\caption[]{a) The order parameter space and the two types of topologically distinct loops: a trivial loop and a vortex. The loops can lie on the surface of (black) or inside (gray) the solid sphere. The loops corresponding to a vortex connect two antipodal points on the surface of the sphere. b) The nontrivial loop corresponding to the vortex shown in the main text.
} 
\label{FIG:2}
\end{figure}

A rotation in 3D can be parameterized by an angle and an axis. Thus, the order-parameter space can be geometrically represented by a solid sphere of radius $\pi$ with antipodal points on the surface identified: the distance from each point to the center of the sphere represents the angle of rotation, while the vector connecting the point to the center gives the axis of rotation. The identification of antipodal points follows from the fact that clockwise and counterclockwise rotations by angle $\pi$ are equivalent. A 1D loop in real space maps onto a 1D loop in the order-parameter space. As seen in Fig.~\ref{FIG:2}a, due to th identification of antipodal points, there are two topologically distinct loops in the order-parameter space: contractible (topologically trivial), and noncontractible (topologically notrivial). The noncontractible loop  corresponds to a nontrivial vortex. For example, the vortex configuration described in the main text corresponds to the following noncontractible loop: a straight line passing through the North pole, the center of the sphere and the South pole (which is identified with the North pole) of Fig.~\ref{FIG:2}b.

\section{calculation of the response functions}

In this section, we present some details regarding the calculation of different response functions. We use the general linear-response relation for the expectation value of different currents:
\beq
\langle J_a^\nu(\rr,t)\rangle =\int_{t', \rr'}R_{ab}^{\nu\eta} (t-t', \rr - \rr')  {\cal A}_b^\eta(t',\rr'), \label{eq:LR}
\eeq
where $R_{ab}^{\nu\eta}$ are the linear-response functions calculated with respect to the uniform state. 
The number of non-zero response functions can be quite small due to  symmetries. 
In any QH system, the relationship between the charge current and the electromagnetic potential $A^\nu_0$ is given by $\langle J^\eta_0\rangle = \sigma^{00}_{xy} \epsilon_{\eta\mu\nu}\partial^\mu A^\nu_0$, where $\sigma^{00}_{xy}$ is just the regular charge Hall conductivity. Comparing with Eq.~(\ref{eq:LR}), this gives
$$ R_{00}^{\nu\eta} (t-t', \rr - \rr') = \sigma^{00}_{xy} \delta(t-t')\delta(\rr-\rr') \epsilon_{\nu\mu\eta}\partial^\mu.$$ 

Other response functions ${ R}^{\nu\eta}_{ab}(t-t')  = - i \langle[J^\nu_a(t),J^\eta_b(t')]\rangle\theta(t-t')$ involve either spin-spin or spin-charge correlators. For the particular model of Ref. \onlinecite{MB2008}, the only other nonvanishing Hall response is in the spin-spin sector, namely $\sigma^{aa}_{xy}$, $a=1\dots 3$. That is due to the symmetric nature of the ordered state: a combination of a spatial translation and a spin rotation is a symmetry. For other less symmetrical systems, such as an arbitrary spin configuration in the model of Ref.~\onlinecite{nagaosa}, the Hall conductivities will have a less structured form.

All conductivities can be computed from the general expression 
\begin{equation}\label{eq:resp}
\sigma^{ab}_{xy} = \frac{e^2}{h}\frac{1}{2 \pi i} \sum_{n,m,{\bf k}}{\frac {(J^x_a)^{nm\bf k} (J^y_b)^{mn\bf k} }{(\epsilon_{n \bf k} - \epsilon_{m \bf k})^2}[n_F(\epsilon_{n \bf k})-n_F(\epsilon_{m \bf k})]}.
\end{equation}
The expression above is applicable to translationally invariant systems where momentum $\bf k$ is a good quantum number. For a unit cell of $M$ sites, the momentum-space Hamiltonian ${\cal H}(\bf k)$ can be generically written as a $2M\times 2M$ matrix. (The factor of two accounts for electron spin.) For each momentum $\bf k$, we can diagonalize this $2M\times 2M$ Hamiltonian and obtain its eigenvalues and eigenvectors: ${\mathcal H}({\bf k})=\epsilon_{m {\bf k}}|m({\bf k})\rangle$, $m=1\dots 2M$. The eigenvalues $\epsilon_{m {\bf k}}$ give the $2M$ energy bands. If we have symmetries, as in the triangular lattice case discussed below, these bands can be degenerate. The matrix elements $(J)^{nm\bf k}\equiv\langle n({\bf k}) |J({\bf k})|m({\bf k})\rangle$, which appear in Eq.~(\ref{eq:resp}), can be constructed explicitly using the eigenvectors $|m({\bf k})\rangle$ and $ |n({\bf k})\rangle$, where $J({\bf k})$ is an appropriate current operator written as a $2M\times 2M$ matrix in the same basis as ${\mathcal H}({\bf k})$. Also,  $n_F(\epsilon)=\left(1+e^{\beta(\epsilon-\mu)}\right)^{-1}$ is the Fermi occupation number at a given chemical potential $\mu$ and inverse temperature $\beta$. Throughout this work, we focus on zero temperature, where $n_F$ is a step function.
 
The only ingredients for computing Eq.~(\ref{eq:resp}) are then the $2M\times 2M$ Hamiltonian ${\mathcal H}(\bf k)$ and the corresponding $2M\times 2M$ charge and spin current operators $J^{x,y}_a({\bf k})$ for $a=0\dots 3$. With these ingredients, one can diagonalize ${\mathcal H}(\bf k)$ to obtain the eigenvalues and eigenvectors, use the eigenvectors to construct the matrix elements of the current operators, perform the sum over $m$ and $n$, and finally integrate the resulting expression over momenta ${\mathbf k}$ in the Brillouin zone.

 \section{triangular lattice}
 
\begin{figure}[ht]
 \includegraphics[width=14cm]{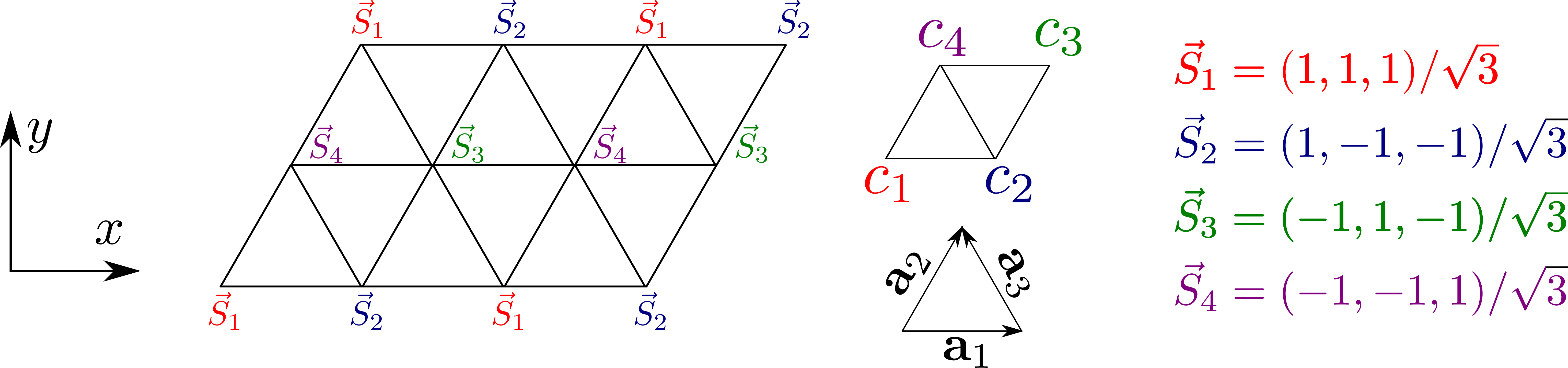}
\caption[]{An explicit chiral configuration of magnetic moments  with tetrahedral ordering on the triangular lattice. For site $i$ in sublattice $a=1\dots4$, ${\mathbf S}_i=\vec{S}_a$ as shown in the figure. The vectors ${\mathbf a}_i$ are the lattice vectors.}
\label{fig:trig} 
\end{figure}

 In this section, we present the details of the calculation of the response functions [Eq. (9) of the main text] on the triangular lattice. Based on the above discussion, our main task is to write ${\mathcal H}({\bf k})$ and $J^{x,y}_a({\bf k})$. We first choose an explicit tetrahedral magnetic ordering represented by the moments $\vec{S}_a$, $a=1\dots 4$ in Fig.~\ref{fig:trig}. With four sublattices and two spin species, we can write the Hamiltonian as an $8\times 8$ matrix in momentum space. As in the main text, we assume an additional Zeeman field $h$ in the $z$ direction. We choose the following basis:
 \[\Psi^\dagger_{\mathbf{k}}=(c^\dagger_{1 \uparrow {\mathbf{k}}},c^\dagger_{1 \downarrow {\mathbf{k}}},c^\dagger_{2 \uparrow {\mathbf{k}}},c^\dagger_{2 \downarrow {\mathbf{k}}},c^\dagger_{3 \uparrow {\mathbf{k}}},c^\dagger_{3 \downarrow {\mathbf{k}}},c^\dagger_{4 \uparrow {\mathbf{k}}},c^\dagger_{4 \downarrow {\mathbf{k}}}),\]
to write the Hamiltonian, i.e., \[H_T=\sum_{\mathbf{k}}\Psi^\dagger_{\mathbf{k}} {\cal H}_T(\mathbf{k})\Psi_{\mathbf{k}},\] where the subscript $T$ indicates the triangular lattice.
We can then write  
\begin{equation}\label{eq:hamil_t}
{\cal H}_T({\mathbf k})=J
\left(\begin{array}{cccc}
\vec{S}_1\cdot{\boldsymbol\sigma} +h \sigma^3&
0 &
 0 &
 0
   \\
0&
\vec{S}_2\cdot{\boldsymbol\sigma} +h \sigma^3&
0&
0
\\
0&
0 &
\vec{S}_3\cdot{\boldsymbol\sigma} +h \sigma^3 &
0\\
0&
0 &
0&
\vec{S}_4\cdot{\boldsymbol\sigma} +h \sigma^3
\end{array} 
\right)+
{\cal E}_T({\mathbf k})\otimes \sigma^0,
\end{equation}
 where
 \begin{equation}
{\cal E}_T({\mathbf k})=
\left(\begin{array}{cccc}
0 &
-2 t \cos ( {\mathbf k}\cdot {\mathbf a}_1)&
-2 t \cos ( {\mathbf k}\cdot {\mathbf a}_3)  &
-2 t \cos ( {\mathbf k}\cdot {\mathbf a}_2)
   \\
-2 t \cos ( {\mathbf k}\cdot {\mathbf a}_1)&
0 &
-2 t \cos ( {\mathbf k}\cdot {\mathbf a}_2)&
-2 t \cos ( {\mathbf k}\cdot {\mathbf a}_3)
\\
-2 t \cos ( {\mathbf k}\cdot {\mathbf a}_3)&
-2 t \cos ( {\mathbf k}\cdot {\mathbf a}_2) &
0 &
-2 t \cos ( {\mathbf k}\cdot {\mathbf a}_1)
\\
 -2 t \cos ( {\mathbf k}\cdot {\mathbf a}_2)&
-2 t \cos ( {\mathbf k}\cdot {\mathbf a}_3)&
-2 t \cos ( {\mathbf k}\cdot {\mathbf a}_1)&
0
\end{array} 
\right).
\end{equation}
 
Charge ($a=0$) and spin ($a=1,2,3$) current operators can then be simply written in the same basis as
\begin{equation}\label{eq:current}
J^x_a({\bf k})=\partial_{k_x}{\cal E}_T({\mathbf k})\otimes \sigma^a,\quad J^y_a({\bf k})=\partial_{k_y}{\cal E}_T({\mathbf k})\otimes \sigma^a.
\end{equation}

 \section{\textit{kagom\'e} lattice}

\begin{figure}[ht]
 \includegraphics[width=14cm]{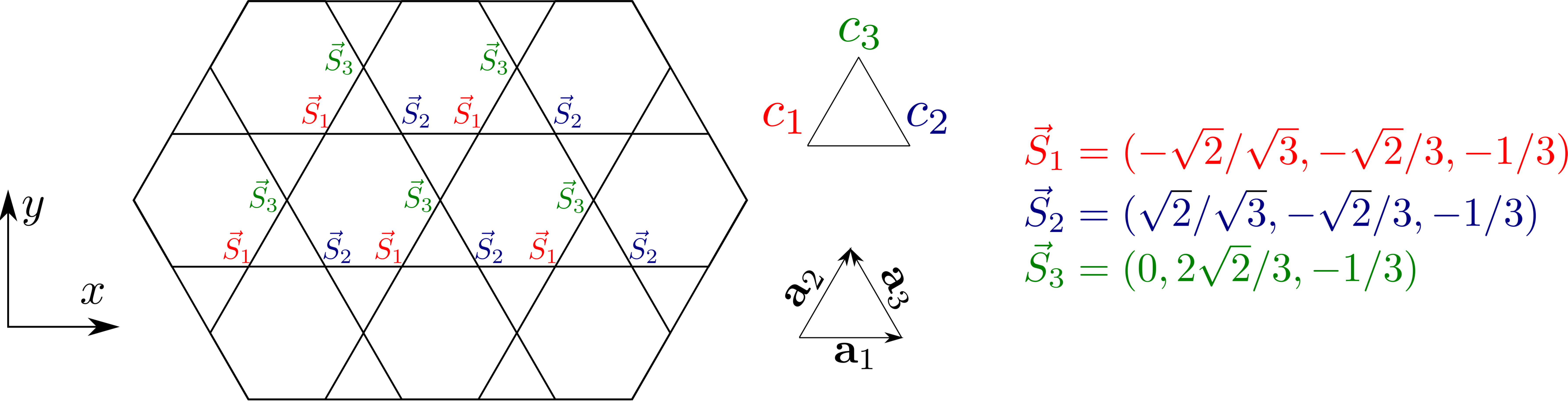}
\caption[]{An explicit chiral configuration of magnetic moments on the \textit{kagom\'e} lattice. For site $i$ in sublattice $a=1\dots3$, ${\mathbf S}_i=\vec{S}_a$ as shown in the figure. The vectors ${\mathbf a}_i$ are the lattice vectors.}
\label{fig:kag} 
\end{figure}

As stated in the main text, the mechanism for fractionalization is generic, and does not depend on the specifics of the triangular lattice model. To demonstrate this fact, we consider another system where noncoplanar magnetic ordering leads to spontaneous integer quantum Hall effect.~\cite{nagaosa} The model is on the \textit{kagom\'e} lattice, i.e., a lattice of corner-sharing triangles, shown in Fig.~\ref{fig:kag}. In Ref.~[\onlinecite{nagaosa}], a limit of large Kondo coupling was assumed, where the problem reduces to one of spinless electrons hopping on the lattice with a magnetic flux $\phi$ ($-2\phi$) inserted in each triangular (hexagonal) plaquette.  We choose the following noncoplanar ordered texture represented in Fig.~\ref{fig:kag}: the projections of the magnetic moments $\vec{S}_a$, $a=1\dots 3$, on the $xy$ plane form $120$-degree ordering on this plane. All the moments, however, are canted down perpendicular to the $xy$ plane, creating a noncoplanar texture. In the limit of large $J$, this setup reduces to the model of Ref.~[\onlinecite{nagaosa}]. For concreteness, we work with the specific moment configurartion shown in Fig.~\ref{fig:kag}, where three moments point to the corners of a tetrahedron, as in the case of the triangular lattice. Other canting angles would give rise to similar physics.  In analogy with the triangular lattice case in the main text, we introduce as an extra tuning parameter in the Hamiltonian a Zeeman field $h$ in the $z$ direction. 

First, we calculate the different response functions $\sigma_{xy}^{0a}$ and $\sigma_{yx}^{0a}$, for $a=0\dots3$, which can lead to the accumulation of charge, when adiabatically turning on a $Z_2$ vortex, via the Laughlin argument. Then, we compute the charge accumulation around the vortex by exact diagonalization in a finite system. We obtain very good agreement between the two methods. Interestingly, in the absence of a Zeeman field, we obtain a charge $1/2$, just as in the traiangular lattice case, when the spin rotation axis lies in the $xy$ plane. 

As before, to calculate the responses, we simply need to write down ${\mathcal H}({\bf k})$ and $J^{x,y}_a({\bf k})$. 

\begin{figure}[ht]
\vspace{4mm}
 \includegraphics[width=8cm]{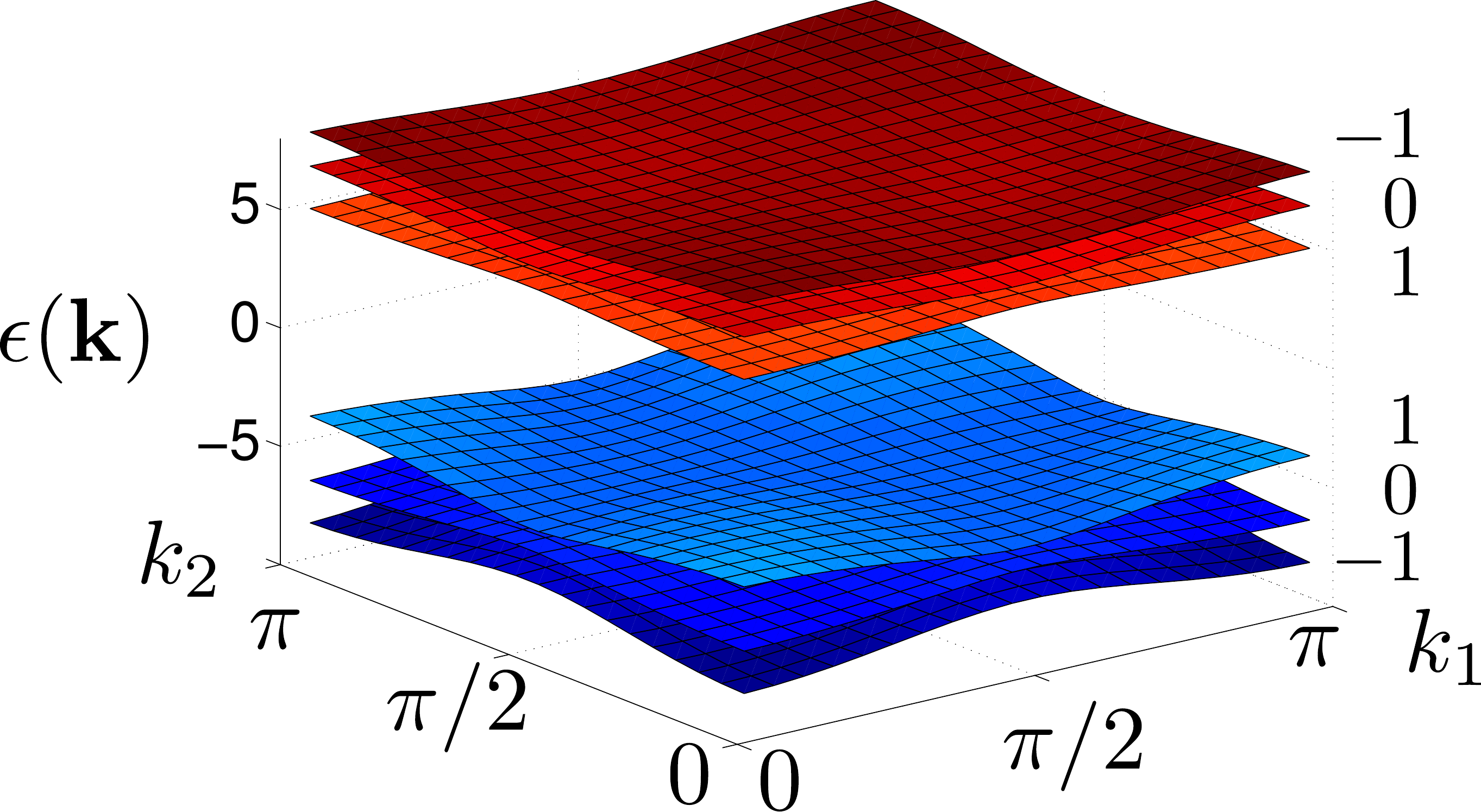}
\caption[]{The band structure of Hamiltonian~(\ref{eq:hamil}) for $J=6$ and $h=0$, where $\mathbf k=k_1 \mathbf {Q}_1+k_2 \mathbf {Q}_2$, and $\mathbf {Q}_1=(1,-1/\sqrt{3})$ and $\mathbf {Q}_2=(0,2/\sqrt{3})$ are reciprocal vectors. The Chern numbers of the bands are shown on the right-hand side. From the Chern numbers, we find that the system exhibits integer quantum Hall effect at $1/6$, $2/6$, $4/6$ and $5/6$ filling fraction.}
\label{fig:band} 
\end{figure}

Defining
\[\Psi^\dagger_{\mathbf{k}}=(c^\dagger_{1 \uparrow {\mathbf{k}}},c^\dagger_{1 \downarrow {\mathbf{k}}},c^\dagger_{2 \uparrow {\mathbf{k}}},c^\dagger_{2 \downarrow {\mathbf{k}}},c^\dagger_{3 \uparrow {\mathbf{k}}},c^\dagger_{3 \downarrow {\mathbf{k}}}),\]
we can write the Hamiltonian as $H_K=\sum_{\mathbf{k}}\Psi^\dagger_{\mathbf{k}} {\mathcal H}_K(\mathbf{k})\Psi_{\mathbf{k}}$, where the $6\times 6$ ${\mathcal H}_K(\mathbf{k})$ matrix is given by
\begin{equation}\label{eq:hamil}
{\cal H}_K({\mathbf k})=J
\left(\begin{array}{ccc}
\vec{S}_1\cdot{\boldsymbol\sigma}+h \sigma^3 &
0 &
 0 
   \\
0&
\vec{S}_2\cdot{\boldsymbol\sigma}+h \sigma^3 &
0
\\
0&
0 &
\vec{S}_3\cdot{\boldsymbol\sigma} +h \sigma^3
\end{array} 
\right)+
{\cal E}_K({\mathbf k})\otimes \sigma^0,
\end{equation}
with $\sigma^0$ representing the $2\times 2$ identity matrix and the $3\times 3$ matrix ${\cal E}_K({\mathbf k})$ given by 
\begin{equation}
{\cal E}_K({\mathbf k})=
\left(\begin{array}{cccc}
0 &
-2 t \cos ( {\mathbf k}\cdot {\mathbf a}_1)&
-2 t \cos ( {\mathbf k}\cdot {\mathbf a}_2)
   \\
-2 t \cos ( {\mathbf k}\cdot {\mathbf a}_1)&
0 &
-2 t \cos ( {\mathbf k}\cdot {\mathbf a}_3)
\\
-2 t \cos ( {\mathbf k}\cdot {\mathbf a}_2)&
-2 t \cos ( {\mathbf k}\cdot {\mathbf a}_3)&
 0
\end{array} 
\right).
\end{equation}
The lattice vectors ${\mathbf a}_i$ are defined in Fig.~\ref{fig:kag}, and the subscript $K$ indicates the \textit{kagom\'e} lattice. The spectrum of the Hamiltonian above consists of $6$ bands, as shown in Fig.~\ref{fig:band}, e.g., for $J=6$ and $h=0$. In the limit of large $J$, the three lower bands are separated from the three upper ones by an energy scale proportional to $J$. Throughout this section, we only consider the $5/6$ filling, i.e., we work at a chemical potential where only the highest energy band is unoccupied. This is one of the several fillings where the system exhibits integer quantum Hall response~\cite{nagaosa}, as seen in Fig.~\ref{fig:band}.
Charge ($a=0$) and spin ($a=1,2,3$) current operators can then be simply written as in Eq.~(\ref{eq:current}), with ${\cal E}_T$ replaced by ${\cal E}_K$.
The response functions for $h=0$ have the following properties:
\[
\sigma_{xy}^{00}=-\sigma_{yx}^{00}=e^2/h,\quad \sigma_{xy}^{01}\approx\sigma_{yx}^{01}\approx 0,\quad\sigma_{xy}^{02}=\sigma_{yx}^{02},\quad\sigma_{xy}^{03}=-\sigma_{yx}^{03}.
\]
As expected all the response functions saturate to fixed values for large $J$, where the model reduces to the model of Ref.~[\onlinecite{nagaosa}].

\begin{figure}[h]
\vspace{4mm}
 \includegraphics[width=15cm]{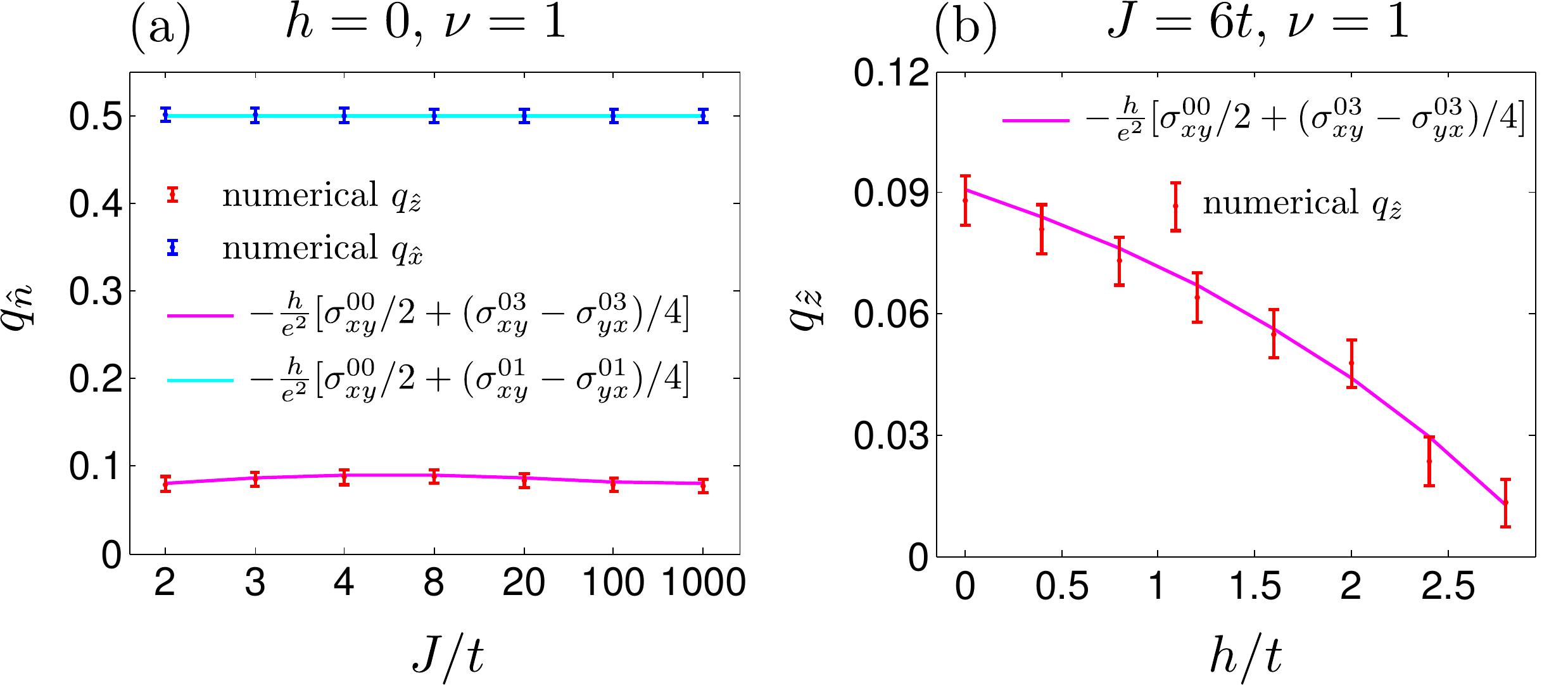}
\caption[]{ a) Vortex charge for the \textit{kagom\'e} lattice with different coupling strengths. Red (blue) dots were obtained by exact diagonalization for a vortex with rotations around the $\hat{z}$ ($\hat{x}$) axis. The solid lines are the analytical predictions from Lauglin's argument argument. Data for $\hat{y}$-axis vortices are very similar to the $\hat{x}$ one and are not shown for clarity. b) Zeeman-field dependence of the vortex charge for the \textit{kagom\'e} lattice. Red dots were obtained by exact diagonalization. The solid line is the result expected from Laughlin's adiabatic argument. The electronic filling is $5/6$ in both cases.}
\label{fig:plot} 
\end{figure}

Following the argument in the main text, the charge trapped by a vortex with the spin rotation axis around the direction $\hat{n}=\hat{x},\hat{y},\hat{z}$ is then given by 
\begin{equation}
q_{\hat{n}} = {h \over e^2}[-\sigma^{00}_{xy}/2 - (\sigma^{0n}_{xy} - \sigma^{0n}_{yx})/4].
\end{equation}
Fig.~\ref{fig:plot}a shows the charge of a vortex for several coupling strengths. The horizontal axis does not have a linear scale so both the variations at small $J$ and the saturation for large $J$ can be displayed.
The results computed from Laughlin's argument show very good agreement with the numerical ones.
Fig. \ref{fig:plot}b shows the charge $q_{\hat z}$ trapped by a vortex as a function of the Zeeman field  $h$ along $\hat{z}$ axis.
Like the triangular lattice case, we find that the charge Hall conductivity $\sigma^{00}_{xy}$ does not change as long as the spectral gap remain open. The off-diagonal responses $\sigma^{03}_{xy}$ on the other hand change as a function of $h$ resulting in a continuously changing, generally irrational, fractional charge.  
The results computed from the Laughlin's argument again show excellent agreement with the ones obtained by numerical diagonalization on finite lattices of size $24 \times 24$.